\documentclass[12pt]{iopart}
\usepackage{pstricks}%
\usepackage{graphicx}
\usepackage{dcolumn}
\usepackage{bm}
\usepackage{amssymb}
\usepackage{iopams}
\usepackage{pifont}
\usepackage{psfrag}
\usepackage[latin2]{inputenc}
\usepackage{oldgerm}
\usepackage{cite}
\usepackage{enumerate}

\newcommand{\bra}[1]{\langle{#1}|}
\newcommand{\ket}[1]{|{#1}\rangle}
\newcommand{\braket}[2]{\langle{#1}|{#2}\rangle}

\newcommand{\BE}{\begin{equation}}
\newcommand{\EE}{\end{equation}}
\newcommand{\I}{{\rm{i}}}
\newcommand{\fig}[1]{figure~\ref{#1}}
\newcommand{\eq}[1]{(\ref{#1})}
\newcommand{\modu}{{\rm{ \;mod\; }}}
\newcommand{\xrightarrow}[2]{\begin{array}{c}
_{#1}\\
\overrightarrow{^{\;\;#2\;\;}}
\end{array}}

\begin{document}

\title[Factorization with Gauss sums:  Entanglement]{Factorization of numbers with Gauss sums: \\III. Algorithms with Entanglement}
\author{S W\"olk$^1$ and W P Schleich$^1$}

\address{$^1$ Institute of Quantum Physics and Center for Integrated Quantum Science and Technology, Ulm University,\\ Albert-Einstein-Allee 11, D-89081 Ulm, Germany 
}%

\ead{sabine.woelk@uni-ulm.de}

\date{\today}

\date{\today}

\begin{abstract}
We propose two algorithms to factor numbers using Gauss sums and entanglement: (i) in a Shor-like algorithm we encode the standard Gauss sum in one of two entangled states and (ii) in an interference algorithm we create a superposition of Gauss sums in the probability amplitudes of two entangled states.These schemes are rather efficient provided that there exists a fast algorithm 
 that can detect a period of a function hidden in its zeros. 

\end{abstract}

\pacs{Valid PACS appear here}

\section{Introduction}

Gauss sums, that are sums whose phases depend quadratically on the summation index, have periodicity properties that make them ideal tools to factor numbers. The crucial role of periodicity in the celebrated Shor algorithm has recently been identified and summarized by  N.~D.~Mermin \cite{Mermin2007}  in the statement ``{\it Quantum mechanics is connected to factoring through periodicity ... and a quantum computer provides an extremely efficient way to find periods}''.

In a series of papers\cite{Woelk2011,Merkel2011} we have analyzed the possibilities  of Gauss sums for factorization offered by their periodicity properties. Although our considerations were confirmed by numerous experiments\cite{Experiments} the schemes proposed so far scale exponentially since they do not envolve entanglement. 
In the present article we propose and investigate two algorithms which connect\cite{doc,suter}  Gauss sum factorization with entanglement. 

Throughout our article, we consider two interacting quantum systems and describe them by two complete sets of states with discrete eigenvalues. We pursue two approaches: (i)  we encode the absolute value of the standard Gauss sum in one of the two quantum states, and (ii) we create an interference of Gauss sums in the probability amplitudes of a quantum state.

Since our first  algorithm is inspired by the one of Shor, we replace the  modular exponentiation $f$ used by Shor by a function $g$ defined by the standard Gauss sum. However, there is a crucial difference between $f$  and $g$: whereas every value of $f$ is assumed in a period only  once \cite{Mermin2007}, the function $g$ takes on the same value several times. In this case, the periodicity is stored in the zeros of a probability distribution. Moreover,  this method is based on a very specific initial state which is unfortunately hard to realize. In order to avoid these complications, we encode in the second approach the Gauss sum in the probability amplitudes of the state rather than in the state itself. In this way we obtain a superposition of Gauss sums.

Our article is organized as followes: in Sec. \ref{chap:like_Shor} we combine the Shor algorithm with Gauss sum factorization by replacing the function $f$ by the appropriately standard Gauss sum $g$. The discussion of this new algorithm leads in Sec. \ref{chap:entanglement} to the idea  of using entanglement to estimate the Gauss sum ${\cal W}_n^{(N)}$, which we then apply to factor numbers.  We conclude in Sec. \ref{sec:summary} by summarizing our results and presenting an outlook.



\section{ Shor algorithm with Gauss sum\label{chap:like_Shor}}

In this section, we discuss a generalization of the Shor algorithm where the absolute value of an appropriately normalized standard Gauss sum replaces the modular exponentiation.
For this purpose, we first analyze the periodicity properties of this function and then suggest an algorithm similar to Shor. Next, we investigate the factorization properties depending on the measurement outcome of the second system. We conclude  with a brief discussion of the similarities and differences between the original Shor algorithm and our alternative proposal.


\subsection{Periodicity properties of normalized standard Gauss sum \label{sec:period_G(l,N)}}

The Shor algorithm  \cite{Shor1994} contains two crucial ingredients:  (i) the {\it mathematical property} that the function
\BE
f(\ell,N)\equiv a^{\ell} \modu N \label{def:f(l,N)}
\EE
exhibits a period $r$, that is $f(\ell,N)=f(\ell+r,N)$,  and (ii) the { \it quantum mechanical property} that the Quantum Fourier Transform (QFT) is able to find the period of a function in an efficient way. 

However, we now show that it is possible to construct an algorithm similar to the one by Shor, by using the periods of other functions which also contain information about the factors of a given number $N$. An example  is 
the function
\BE
g(\ell,N)\equiv \frac{1}{N}|G(\ell,N)|^2\label{def:g(l,N)} 
\EE
 expressed in terms of the standard Gauss sum \cite{number_theory,schleich:2005:primes}
\BE
G(\ell,N)\equiv \sum\limits_{m=0}^{N-1}\exp\left[2\pi \I m^2 \frac{\ell}{N}\right].\label{stan_Gauss}
\EE

The properties of $G$ provide us with the explicit form
\BE
g={\rm{gcd}}(\ell,N)
\EE
that is the function $g$ is determined by  the greatest common divisor  (gcd) of $N$ and $\ell$.

We now analyze the periodicity properties of $g$ for the two cases: (i)  $N$  consists of two, or (ii) more than two prime factors.

\subsubsection{$N$ consists of two prime factors}

\begin{figure}[ht!]
\begin{center}
\includegraphics[width=0.7\columnwidth]{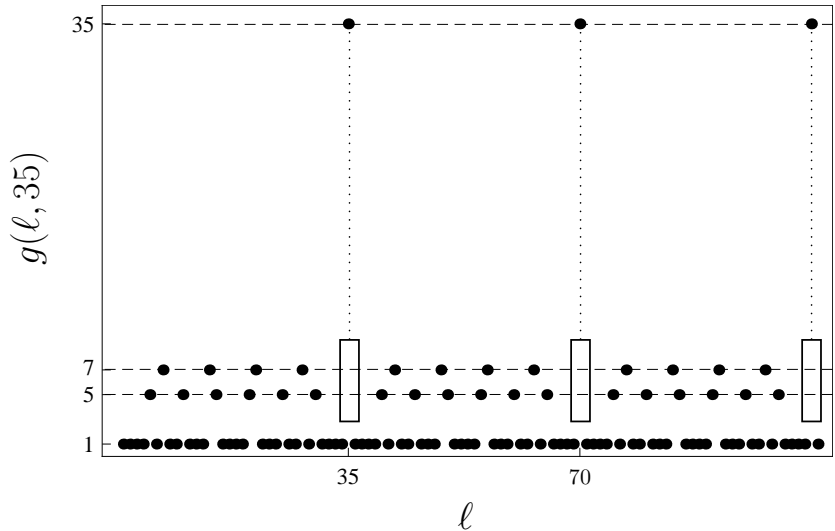}
\end{center}
\caption[Periodicity properties of $g= g(\ell,N)$]{Periodicity properties of the function $g= g(\ell,N)$ defined by \eq{def:g(l,N)} for the example $N=35=5 \cdot 7$. This function contains one perfect period $r=35$, given by the number $N=35$, and two imperfect periods $\tilde r=5,7$ determined by the factors of $N$. At multiples $k$ of a factor  $p=5$ or $7$ the function $g$ is given by the factor itself. However,  if $\ell$ is also a multiple of  $N=35$ as marked by rectangles, the periodicity relation $g(\ell,N)= g(\ell+\tilde r,N)$ does not hold anymore. 
}
\label{fig:G_35}
\end{figure}

If $N$ contains only the two prime factors $p$ and $q$, the explicit value of the function $g= g(\ell,N)$ is given by
\BE
g(\ell,N)=\left\{\begin{array}{ll}
N & {\rm{if \;}}\ell = k \cdot N\\ 
p & {\rm{if \;}}\ell = k \cdot p\\ 
q & {\rm{if \;}}\ell = k \cdot q\\ 
1 & {\rm{else}}
\end{array}\right. .
\EE

As a consequence, $g$ shows one perfect period $r=N $, where the identity $g(\ell,N)=g(\ell+r,N) $ is valid for all arguments $\ell$, and two imperfect periods $\tilde r=p,q$, where  $g(\ell,N)=g(\ell+\tilde r,N) $ is valid for allmost all arguments $\ell$. This behavior of $g$ is  displayed in \fig{fig:G_35} for the example $N=35$. Indeed, every argument $\ell$ of $g$ which is a multiple of a factor $p$ leads to a  value of $g$  equal to this factor. However, for arguments $\ell=k \cdot N$ which are also multiples of $N$ the function $g$  yields $g(k \cdot N,N)=N$, and therefore the periodicity relation $g(\ell,N)= g(\ell+k\cdot p,N)$ does not hold true for these arguments.

Furthermore,  the imperfect periods  given by the factors of $N$ interrupt each other.  For example, all arguments $\ell$ which are multiples of $q$ do not  satisfy the periodicity relation for the imperfect period $\tilde r=p$ because the greatest common divisor of $\ell=sq$ and $N$ is $q$ (if $s\neq k\cdot p$) and therefore
\BE
g(s\cdot q,N)= q.
\EE
However, the argument $\ell=s\cdot q+k\cdot p$ shares in general no factor with $N$. Therefore, we obtain
\BE
g(s\cdot q+k \cdot p,N)= 1.
\EE


\subsubsection{$N$ consists of three or more prime factors}

\begin{figure}[ht!]
\begin{center}
\includegraphics[width=0.7\columnwidth]{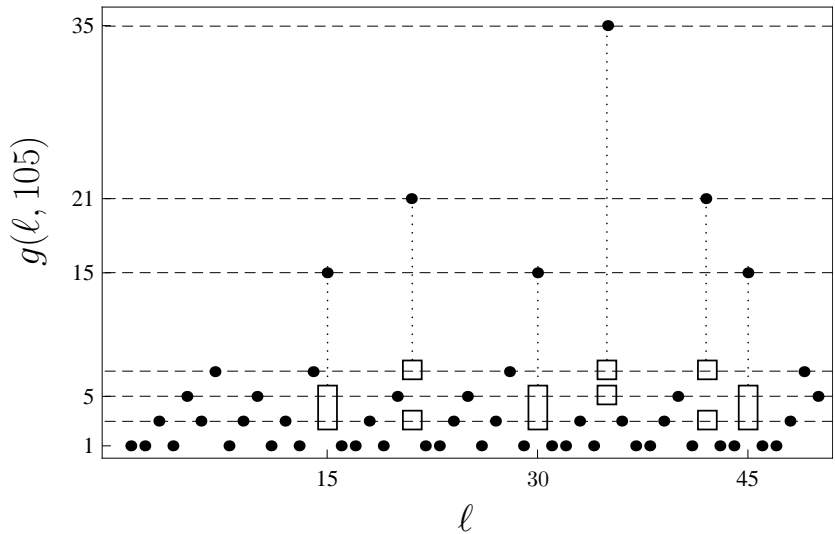}
\end{center}
\caption[Imperfection of the periodicity of $g(\ell,N)$]{For numbers $N$  with more than two factors such as $N=105=3 \cdot 5\cdot 7$, there exist several imperfections of the periodic behavior of $g$ which stands out most clearly for multiples of factors such as $p=5$. Whenever the argument $\ell$ is a multiple of a product of two or more factors, such as $\ell=30=2 \cdot 3\cdot 5$, the signal is enhanced, and the periodicity relation is not valid at these arguments.
}
\label{fig:G_105}
\end{figure}

If $N$ consists of more than two prime factors, such as $N=105=3\cdot 5\cdot 7$, then the signal $g$ is more complicated, as shown in \fig{fig:G_105}. Here, the imperfect periodicity at multiples of factors  is interrupted for every  $\ell$, which shares more than one prime factor with $N$, such as $\ell=30=2 \cdot 3 \cdot5$. However, these arguments $\ell$ form a new imperfect period  given by ${\rm{gcd}}(\ell,N)$ which in our example reads ${\rm{gcd}}(30,105)=15$. This new imperfect period again contains information about the factors of $N$.

In the following sections, we will not distinguish between perfect periods and imperfect periods anymore, since the imperfection of the imperfect periods do not influence our proposed algorithms, as long as $g(\ell,N)= g(\ell+\tilde r,N)$ is valid for most arguments $\ell$.


\subsection{Outline of algorithm\label{sec:algorithm_like_Shor}}

In the present section we introduce an algorithm which combines Gauss sums and entanglement and is constructed in complete analogy to the Shor algorithm. For the sake of simplicity we concentrate on numbers $N$ with only two prime factors $p$ and $q$.

Similar to Shor, we start with the entangled state
\BE
\ket{\Psi}_{A,B}\equiv\frac{1}{\sqrt{2^{Q}}} \sum\limits_{\ell=0}^{2^Q-1}\ket{\ell}_A\ket{g(\ell,N)}_B,\label{eq:like_shor_start}
\EE
of two systems $A$ and $B$. However, in contrast to Shor we encode in system $B$ the function $g$  defined by \eq{def:g(l,N)}  rather than $f$ given by \eq{def:f(l,N)}. 
The dimension of system $A$ is chosen to be $2^Q$ because we want to realize this system with qbits. We will give a condition for the magnitude of $Q$ in the next section.

In the second step, we perform a measurement on system $B$. For an integer  $N=p\cdot q$ consisting only of the two prime factors $p$ and $q$ there exist three distinct   measurement outcomes: 
\begin{enumerate}[(i)]
\item   the number $N$ to be factored, that is $g(\ell,N)=N$,
\item a factor  of $N$, that is $g(\ell,N)=p $ or $g(\ell,N)=q $, 
\item  unity, that is $g(\ell,N)=1$.
\end{enumerate}

In case (i), the state of system $A$ after the measurement of $B$ reads
\BE
\ket{\psi^{(N)}}_{A} \equiv {\cal N}^{(N)} \sum\limits_{k=0}^{M_N-1}\ket{k\cdot N}_{A}
\EE
where the normalization constant $ {\cal N}^{(N)} \equiv M_N^{-1/2}$ is given by $M_N\equiv[2^Q/N]$, and $[x]$ denotes the smallest integer which is larger than $x$. 

The state $\ket{\psi^{(N)}}_{A}$ shows  a periodicity with period $N$, which does not help us to factor the number $N$. Therefore, we will have to repeat the first two steps of our algorithm until the measurement outcome differs from $N$. Fortunately, case (i) occurs only with the probability ${\cal P}_B^{(N)}\approx1/N$ and is therefore not very likely.

In case (ii), system $A$ is in a superposition of all number states $\ket{\ell}_A$, which are multiples of the factor $p$ of $N$, but not of $N$ itself giving rise to
\BE
\ket{\psi^{(p)}}_A\equiv {\cal N}^{(p)}\left(\sum\limits_{k=0}^{M_p-1}\ket{k\cdot p}_A-\sum\limits_{k=0}^{M_N-1}\ket{k\cdot N}_A\right) \label{def:psi_p_A}
\EE
where $ {\cal N}^{(p)} \equiv (M_p-M_N)^{-1/2}$ with $M_p\equiv[2^Q/p]$. 

\begin{figure}[ht]
\begin{center}
\includegraphics[width=0.65\columnwidth]{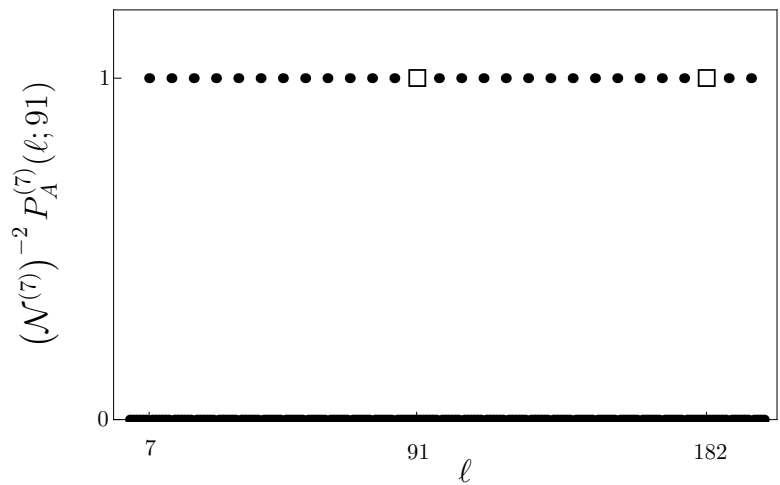}
\end{center}
\caption[Probability distribution $P_A^{(p)}(\ell;91)$ of $\ell$ for the state $\ket{\psi^{(7)}}_A$]{Probability distribution  $P_A^{(7)}(\ell;91)$ to find $\ket{\ell}_A$  in the state $\ket{\psi^{(7)}}_A$ for the example $N=91=7\cdot 13$ and $p=7$ in the range $1\leq \ell \leq 200$. The probability is non-vanishing only for multiples of $7$. However, multiples of $7$ which are also multiples of $N=91$ have a vanishing probability. As a consequence, the period of $\ket{\psi^{(7)}}_A$ is imperfect. 
}
\label{fig:psi_a}
\end{figure}

 In this state,  as depicted in \fig{fig:psi_a} for the example $N=91$ and $p=7$, only multiples of $p$ appear with a non-zero probability 
\BE
P_A^{(p)}(\ell;N) \equiv \left|_A\braket{p}{\psi^{(p)}}_A\right|^2
\EE
which leads to a clearly visible periodicity. However, the periodicity is imperfect at arguments $\ell=k\cdot N$, which are multiples of $N$.  We emphasize that this case occurs with the probability ${\cal P}_B^{(p)}\approx1/p-1/N$ and is therefore more likely than (i).

In the third case, system $A$ contains all numbers $\ell$ which are not multiples of one of the factors of $N$. As a consequence, the state reads
\begin{eqnarray}
\fl \ket{\psi^{(1)}}_A\equiv& {\cal N}^{(1)} \left(\sum\limits_{\ell=0}^{2^Q-1}\ket{\ell}_A - \sum\limits_{k=0}^{M_p-1}\ket{k\cdot p}_A \right.  \left. -\sum\limits_{k=0}^{M_q-1}\ket{k \cdot q}_A+\sum\limits_{k=0}^{M_N-1}\ket{k\cdot N}_A\right)\label{def:psi_3}
\end{eqnarray}
with $ {\cal N}^{(1)} \equiv (2^Q-M_p-M_q+M_N)^{-1/2}$ and $M_q\equiv[2^Q/q]$. Since all multiples of $N$ are  contained in the second  as well as in the  third sum we have subtracted them twice. Therefore, we have to add them once  again. 

The probability for this case is given by
\BE 
{\cal P}_B^{(1)}\approx 1-\frac{1}{p}-\frac{1}{q}+\frac{1}{N}=\frac{N-p-q+1}{N}
\EE
and takes on the smallest value around $1/2$ for $p=2$ and $q=N/2$, but tends to unity for prime factors $p\approx q\approx \sqrt{N}$.

\begin{figure}[ht]
\begin{center}
\includegraphics[width=0.65\columnwidth]{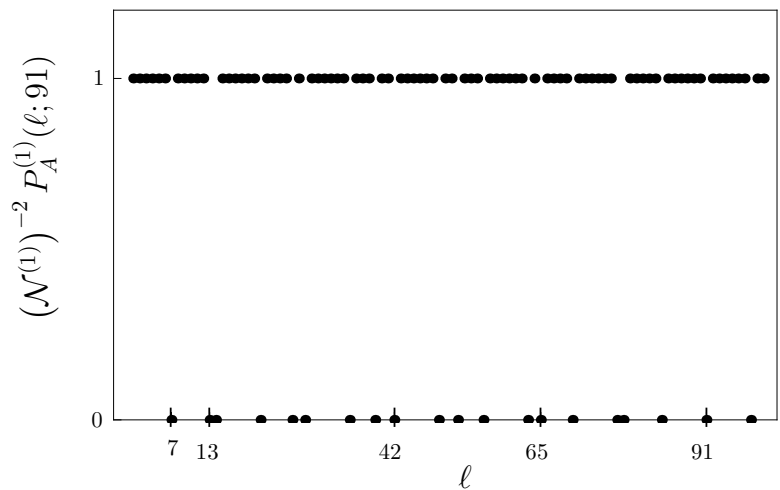}
\end{center}
\caption[Probability distribution of $\ell$ for the state $\ket{\psi^{(1)}}_A$]{Probability distribution  $P_A^{(1)}(\ell;91)$ to find $\ket{\ell}_A$  in the state $\ket{\psi^{(1)}}_A$ for the example $N=91=7\cdot 13$ in the range $1\leq \ell  \leq 100$. The probability vanishes for all multiples of the factors  $7$ and $13$. Here, periodicity  even exists for multiples of $N=91$. The probability is equal to $\left({\cal N}^{(1)}\right)^{-2}$ for all other arguments.
}
\label{fig:psi_b}
\end{figure}

As shown in \fig{fig:psi_b}, the state $\ket{\psi^{(1)}}_A$ exhibits perfect periodicity but every multiple of $p$ and $q$ has zero probability, whereas all other numbers are equally weighted. As a consequence, in this case the information about the factors is encoded in the ``holes''.

\subsection{Analysis of periodicity}

In the preceding section, we have found that  the function $g= g(\ell,N)$ in its dependence on $\ell$ exhibits periods which contain information about the factors of $N$, but in some cases these periods are imperfect.
We now analyze if it is possible to extract information about the periodicity of $g$  with the help of a QFT defined by
\BE
\hat U_{QFT}\ket{\ell}\equiv \frac{1}{\sqrt{2^Q}}\sum\limits_{m=0}^{2^Q-1}\exp\left[2\pi {\I} \frac{ m \ell}{2^Q}\right]\ket{m}.\label{def:Fourier}
\EE
This procedure is analogous to the Shor algorithm. We distinguish two cases for the state of system $A$.

\subsubsection{State of $A$ contains only multiples of $p$\label{sec:one=kp}}

The QFT  transforms the state $\ket{\psi^{(p)}}_A$ given by \eq{def:psi_p_A} into
\BE
\fl\hat U_{QFT}\ket{\psi^{(p)}}_A= \frac{{\cal N}^{(p)}}{\sqrt{2^Q}}\sum\limits_{m=0}^{2^Q-1}
\left[ F\left(\frac{pm}{2^Q};M_p\right) -F\left(\frac{Nm}{2^Q};M_N\right)
\right] \ket{m}_A\label{eq:ls_final_faktor}
\EE
with the definition 
\BE
F(\alpha;M)\equiv \sum\limits_{k=0}^{M-1}\exp\left[2\pi\I k\; \alpha \right]\label{def:F(n,m,Q)2}.
\EE

As shown in \ref{chap:investigation_S}
the sum $F(pm/2^Q;M_p)$ leads to sharp peaks in the probability distribution
\BE
\tilde P_A^{(p)}(m;N)\equiv \left|_A\bra{p}\hat U_{QFT}\ket{\psi^{(p)}}_A\right|^2
\EE
 for $m=m_p$ with
\BE
m_p\equiv j \frac{2^Q}{p}+\delta_j.
\EE
 These peaks will give us information about the factor $p$. 

Unfortunately, the  sum $F(Nm/2^Q;M_N)$  also leads to sharp peaks located at $m_N\equiv j 2^Q/N+\delta_j$.  As a consequence, we need to calculate the probability $\tilde{\cal P}_A^{(p,p)}$  to find any $m_p$ and compare it to the probability $\tilde {\cal P}_A^{(p,N)}$ to measure any $m_N$.

\begin{figure}[ht]
\begin{center}
\includegraphics[width=0.65\columnwidth]{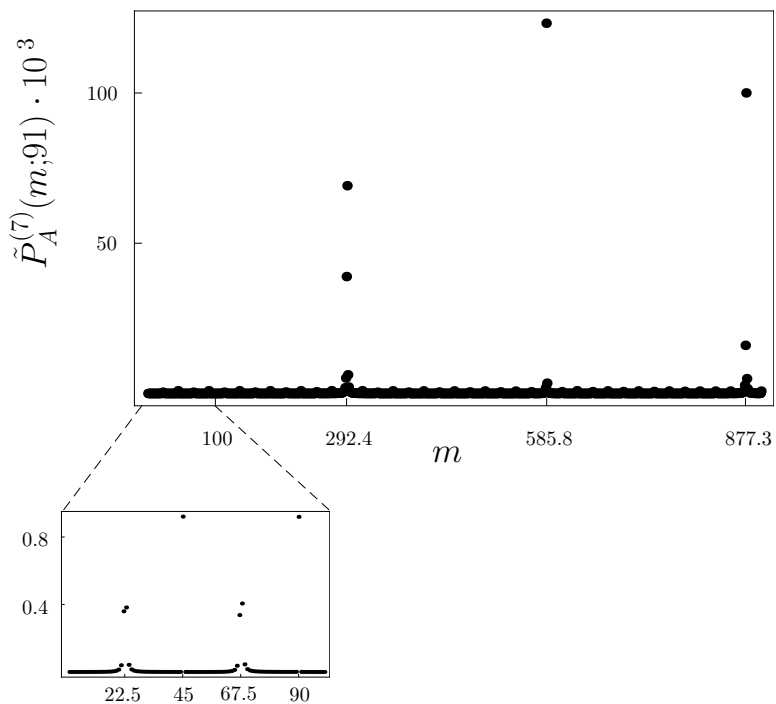}
\end{center}
\caption [Probability distribution $\tilde P_A^{(p)}(m;91)$  if the result of the measurement of system $B$ was equal to $p$]{Probability distribution $\tilde P_A^{(7)}(m;91)$   if the  measurement of system $B$ resulted in the factor  $\ket{7}_B$ shown for the example of $N=91=7 \cdot 13$. Here, we have used a system of $Q=11$ qbits. Clearly visible are peaks at multiples of $2^{11}/7\approx 292.4$. The inset at the bottom magnifies the distribution in the range $1\leq n\leq 100$, where peaks at multiples of $2^{11}/91\approx 22.5$ exist. However, they are approximately $13^2$ times smaller than those at multiples of $292.4$ 
}
\label{fig:lS_faktor}
\end{figure}

In \ref{app:P_p}, we obtain the estimates 
\BE
\tilde{P}_A^{(p)}(m_p;N) >0.4 \frac{N-p}{Np}
\EE
and
\BE
\tilde{P}_A^{(p)}(m_N;N) > 0.4 \frac{p}{N(N-p)}.
\EE
As a consequence, we find that peaks at $m_p$ are  enhanced compared to peaks at $m_N$ by the factor $(N-p)^2/p^2 \approx q^2$. This fact is clearly visible in \fig{fig:lS_faktor} for the example $N=91=7 \cdot 13$.  The large frame shows the peaks located at $m_p$. In the inset we magnify the probability distribution in the range $1\leq m\leq 100$. Here,  also peaks at $m_N$ exist. However, they  are approximately $13^2$ times smaller. Furthermore, as verified in \ref{app:P_p} the total probability $\tilde{\cal P}_A^{(p,p)}$ to find any $m_p$ tends to $0.4$ whereas the probability $\tilde{\cal P}_A^{(p,N)}$ to find any $m_N$ approaches  zero.

At last, we have to analyze the width of the probability distribution for estimating the minimal dimension $2^Q$ for an unique estimation of $p$ from $m_p$. Here, we follow the considerations of N.~D.~Mermin\cite{Mermin2007}.

The measurement result $m$ is within $1/2$ of $j \cdot 2^Q /p$ and therefore
\BE
\left|\frac{m}{2^Q}-\frac{j}{p}\right|\leq \frac{1}{2^{Q+1}}.
\EE
Does there exist another combination $j'/p'\neq j/p$ which lies within this range of $m/2^Q$? The distance between these two pairs of numbers can be approximated by
\BE
\left|\frac{j}{p}-\frac{j'}{p'}\right|= \left|\frac{jp'-j'p}{pp'}\right|\geq \frac{1}{pp'}\geq\frac{1}{N^2}.
\EE
If both combinations would lie whithin $1/2^{Q+1}$ of $m/2^Q$ then their distance would be smaller than,  or equal to $1/2^Q$. 

\begin{figure}[ht!]
\begin{center}
\includegraphics[width=0.6\columnwidth]{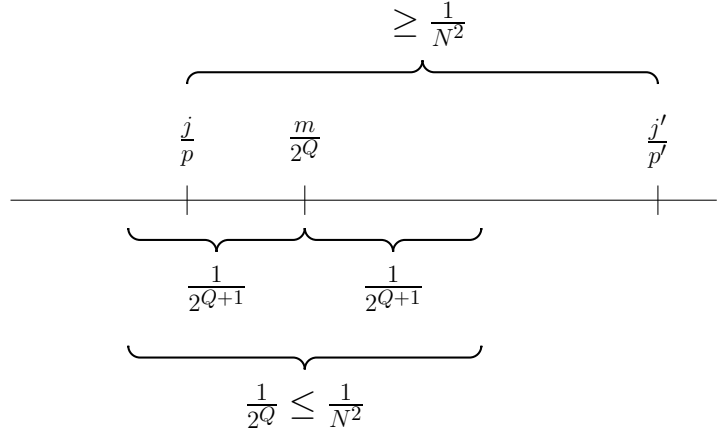}
\caption[Graphical demonstration of uniqueness of the $j/p$]{Graphical demonstration of the uniqueness of $j/p$. The ratio $j/p$ must be within $1/2^{Q+1}$ of $m/2^Q$. Every other combination $j'/p'$ with $j'/p' \neq j/p$ must  differ at least $1/N^2$ from $j/p$. Therefore, if $N^2 < 2^Q$ then $j'/p'$ cannot be within $1/2^{Q+1}$ of $m/2^Q$, too.}
\label{fig:eindeutigkeit}
\end{center}
\end{figure}

Therefore, if $2^Q> N^2$, then their exists only one combination $j\cdot 2^Q/p$ which lies within $1/2$ of $m$. For a graphical representation of this statement, we refer to  \fig{fig:eindeutigkeit}.

We conclude by emphasizing that we can extract in $40\%$ of the measurements  the factor $p$ of $N$. Furthermore, the imperfection of the periodicity does not influence  the ability of the QFT to find the period $p$.


\subsubsection{State of $A$ does not contain any multiples of $p$ or $q$\label{sec:one_neq_kp}}

When we perform the QFT on the state $\ket{\psi^{(1)}}_A$ given by \eq{def:psi_3} we arrive at
\begin{eqnarray}
\fl \hat U_{QFT}\ket{\psi^{(1)}}_A&=& \frac{{\cal N}^{(1)}}{\sqrt{2^Q}} \sum\limits_{m=0}^{2^Q-1}\left[F\left(\frac{m}{2^Q};2^Q\right)-F\left(\frac{pm}{2^Q};M_p\right)\right. \left.-F\left(\frac{qm}{2^Q};M_q\right)\right. \nonumber \\ && \left.+F\left(\frac{Nm}{2^Q};M_N\right)\right]\ket{m}_A\label{eq:QFT_psi_1_A}
\end{eqnarray}
where we have recalled the definition of $F$ from \eq{def:F(n,m,Q)2}.

\begin{figure}[ht]
\begin{center}
\includegraphics[width=0.9\columnwidth]{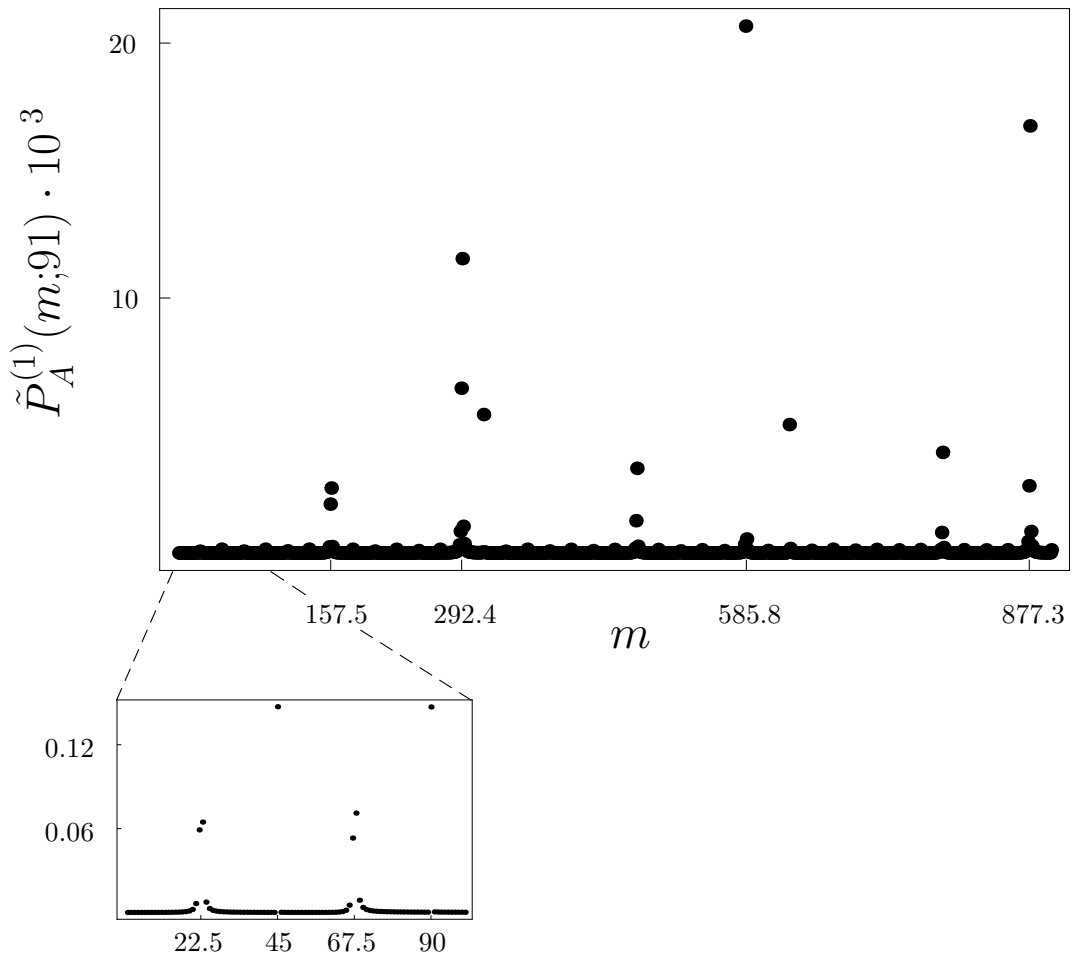}
\end{center}
\caption[Probability distribution $\tilde P_A^{(1)}(m;91)$  of system $A$ conditioned on  the measurement of system $B$ with $g(\ell,N=1)$ and the QFT]{Probability distribution $\tilde P_A^{(1)}(m;91)$  of system $A$ conditioned on  the measurement of system $B$ with the result $\ket{1}_B$ and the QFT for the example $N=91=7 \cdot 13$. Here, we have used a system of $11$ qbits. Clearly visible are peaks at multiples of $2^{11}/13\approx 157.5$ and $2^{11}/7\approx 292.4$. The inset  magnifies the range $1\leq n\leq 100$ where also peaks at multiples of $2^{11}/91\approx 22.5$ exist. However, they are approximately $13^2$ times smaller than the peaks at multiples of $292.4$. A comparison with \fig{fig:lS_faktor} shows that all peaks are smaller than in the case (ii).
}
\label{fig:lS_kfaktor}
\end{figure}

As estimated in \ref{app:tilde_P_2}, the peaks at $m_p$ and $m_q\equiv j\cdot 2^Q/q+\delta_j$ are  approximately $q$ times,  or $p$ times higher than peaks at $m_N$. However, they are smaller in comparison to case (ii), where the measured value of system $B$ was equal to $p$.  All these aspects are visible in \fig{fig:lS_kfaktor} for the example $N=91=7 \cdot 13$. 

The total probability to find any $m_p$ or $m_q$ is given by
\BE
\tilde {\cal P}_A^{(1,p \;{\rm{ or }}\;q)}= p \tilde P_A^{(1)}(m_p;N)+  q \tilde P_A^{(1)}(m_q;N)=\frac{Nq+Np+q+p-4N}{N(N-q-p+1)},
\EE
as calculated in  \ref{app:tilde_P_2}. This probability tends to zero for large $N$ with two prime factors which are of the order of $\sqrt{N}$. As a consequence, it is not useful to try to find the period of $\ket{\psi^{(1)}}_A$ with a QFT.

The periodicity of $\ket{\psi^{(1)}}_A$ is perfect, since the states corresponding to integer multiples of $p$ and $q$ are missing. Moreover,there exist many points with the same value. In contrast, in the Shor algorithm, every integer $\ell$ in the range $0\leq \ell \leq r-1$ yields \cite{Mermin2007} a different outcome  of $f$, and therefore, we are able to find the period $r$ with the help of a QFT. In our scheme there exist approximately $p$ numbers $\ell$ in the range $0 \leq \ell \leq p-1$ which lead to the same value of $g$. As a consequence, it is not possible anymore to find the period $p$ with the help of a QFT. 
Nevertheless, the state $\ket{\psi^{(1)}}_{A}$ is  remarkable, because the information about the factors of $N$ is still endcoded in the periodicity. Since the QFT is not a good tool to find this period, we need to develop another instrument  which  extracts the information about the periodicity of $\ket{\psi^{(1)}}_A$ in an efficient way.

\subsection{Discussion\label{sec:disc_like_Shor}}

In this section, we have analyzed a factorization algorithm  constructed in complete analogy to the one by Shor but with the function $g$ given by \eq{def:g(l,N)} instead of $f$ determined by \eq{def:f(l,N)}. Our approach combines the periodicity properties of Gauss sums with the QFT.  
Although we have found some rather encouraging results there are problems with this approach. Indeed, it is not possible to measure a period with the help of a QFT if there exists a large amount of  arguments $\ell$ within one period were the function assumes  the same  value. As a consequence, the problem of our scheme is not the imperfection of the periodicity of $g$ but the large number of arguments $\ell$ with $g(\ell,N)=1$.

Furthermore, we emphasize that the QFT for the case (ii) is not necessary. A measurement of the state $\ket{\psi^{(p)}}_A$, given by \eq{def:psi_p_A}, itself will also give us the information about the period $p$. In contrast, the  Shor algorithm relies on the state
\BE
\ket{\phi}\equiv \frac{1}{\sqrt{M}}\sum_{m=0}^{M-1}\ket{\ell_0+mr}_{A}
\EE
which contains not only the period $r$, but also the unknown variable $\ell_0$. Since it is not possible to extract  $r$ from an argument $\ell_0+mr$,  the QFT is essential in the Shor algorithm. Furthermore, several runs of the Shor algorithm lead to different numbers $\ell_0$ 

In summary, the modular exponentiation is not only special, because its period contains information about the factors of $N$, but also because every  value appears only once \cite{Mermin2007} in a period. Moreover,  we suspect that there may  exist quantum algorithms, which find periods of a function in an efficient way despite the fact that nearly all arguments $\ell$ in one period assume the same value. This may be achieved for example by a comparison of this periodic function with a function were all arguments assume the same value. Interferometry could archieve such a task. Here, two possibilities offer themselves: i) We can couple the systems $A$ und $B$ to an
ancilla system and prepare the superposition by a Hadamard transform. However, in this
case we are confronted with a probabilistic approach in the spirit of quantum state
engineering \cite{Vogel}. ii) In order to avoid small probabilities for the desired
measurement outcomes we rather pursue an idea based on adiabatic passage, a technique
that has already been used successfully in many situations of quantum optics to
synthesize quantum states \cite{Parkins}.


\section{Algorithm based on a superposition of Gauss sums\label{chap:entanglement}}

In Sec. \ref{chap:like_Shor} we have discussed an algorithm inspired by the one of Shor \cite{Shor1994} which uses the  Gauss sum $g$ instead of  the modular exponentiation. However, we did not explain how to create the initial state
$\ket{\Psi}_{A,B}$ defined by \eq{eq:like_shor_start}. 
Indeed, this task is quite complicated, because in general it is not possible to display and add in an exact way complex numbers of the form ${\rm{e}}^{\I \varphi}$  with  a finite amount of qbits. Therefore, we pursue in this section another approach where we encode  $g$ not in the states $\ket{\ell}_B$ of  system $B$  but in their probability amplitudes.  

\subsection{Central idea}

Encoding a Gauss sum in a probability amplitude of a quantum state was already done experimentally \cite{Experiments} for the truncated Gauss sum 
\BE
{\cal A}^{(M)}_N(\ell)\equiv \frac{1}{M+1}\sum\limits_{m=0}^M\exp\left[2\pi \I m^2\frac{N}{\ell}\right].
\EE
Moreover, the number $M+1$ of terms  of  ${\cal A}_N^{(M)}$ grows only polynomial with the number of digits of $N$, whereas
for the standard Gauss sum $G$, defined by \eq{stan_Gauss}, it increases exponentially. Furthermore, we  want to estimate $G$ for several $\ell$ in parallel. For this reason, it is not useful to realize  $G$ experimentally by a pulse train, or in a ladder system or by interferometry as proposed in \cite{Merkel2011}.

On the other hand, $G$ has two major advantages compared to ${\cal A}_N^{(M)}$: (i)  $\left|G\right|^2$ shows an enhanced signal not only at arguments $\ell=p$ but also at integer multiples  of factors, and (ii) the signal at multiples of a factor is enhanced by the factor itself and not only by $\sqrt{2}$. As a consequence, for the state
\BE
\ket{\psi}\equiv \mathcal{N}\sum\limits_{\ell=0}^{N-1}G(\ell,N)\ket{\ell}\label{eq:traumzustand}
\EE 
the probability to find the state $\ket{\ell}$ with $\ell$ being any multiple of a factor $p$ is $p$ times higher than finding $\ket{\ell\neq k\cdot p}$. The amount of arguments $\ell$, which are not multiples of $p$, is approximatly $p$ times higher than the amount of multiples of $p$. As a consequence, the product of the probability of finding $\ell=k\cdot p$  times the amount of numbers $\ell=k\cdot p$ with $0\leq \ell \leq N-1$ is approximately equal to  the product of the probability of finding $\ell\neq k\cdot p$ times the amount of numbers $\ell\neq k\cdot p$ with $0\leq \ell \leq N-1$. Therefore, the probability to find any multiple of a factor is around $50\%$.

As a result, we have found  a fast factorization algorithm provided we are able to prepare the state $\ket{\psi}$ defined in \eq{eq:traumzustand} in an efficient way. Unfortunately, this is not an easy task. On the other hand, we can use entanglement to calculate the sum
\BE
{\cal W}_n^{(N)}(\ell)\equiv \frac{1}{N}\sum\limits_{m=0}^{N-1}\exp\left[2\pi\I \; (m^2\frac{\ell}{N}+m \frac{n}{N})\right],\label{def:cal_W}
\EE
which is  very close to the standard Gauss sum $G$  and shows similar properties. We therefore now propose an algorithm based on ${\cal W}_n^{(N)}$.

\subsection{Algorithm\label{sec:algorithm}}

The idea of our algorithm is that system $A$ is in a state of superposition of all trial factors $\ell$ and the summation in $m$ in the Gauss sum ${\cal W}_n^{(N)}$ is realized by a superposition of system $B$. Therefore, we start from the product state 
\BE
\ket{\Psi_0}_{A,B}\equiv \frac{1}{N}\sum\limits_{\ell,m=0}^{N-1}\ket{\ell}_A\ket{m}_B\label{def:Psi_0}
\EE
where the dimensions of the systems $A$ and $B$ are equal to the number $N$ to be factored.

Next, we produce phase factors of the form $\exp\left[2\pi \I m^2 \ell/N\right]$which appear in the Gauss sum ${\cal W}_n^{(N)}$ by realizing the unitary transformation
\BE
\hat U_{ph} \equiv \exp\left[2\pi \I \; \hat n_B^2\;\frac{\hat n_A}{N}\right]\label{def:U_ph}.
\EE
Here, $\hat n_j$ denotes the number operator of the system $j=A,B$ and $N$  defines the periodicity of the phase. 

The operator $\hat U_{ph}$  entangles the two systems, and the state $\ket{\Psi_1}_{A,B}$ of the combined system is now given by
\BE
\ket{\Psi_1}_{A,B}\equiv \hat U_{ph}\ket{\Psi_0}_{A,B}= \frac{1}{N}\sum\limits_{\ell,m=0}^{N-1}\exp\left[2\pi \I \;  m^2\frac{\ell}{N}\right]\;\ket{\ell}_A\ket{m}_B.
\EE

We emphasize,  that the information about the Gauss  sum is not stored in a single  system, but in the phase relations between the two systems. Therefore, tracing out one system  and applying a number state measurement on the other, or measuring the number states of both systems would not help us to estimate the Gauss sums ${\cal W}_n^{(N)}$. It would only show that all trial factors have equal weight. As a consequence,  we have to perform local operations on the individual systems, which do not destroy the information inherent in the phase relations but help us to read out the Gauss sum.

Therefore, we perform as a second step a QFT, as defined in \eq{def:Fourier} on system $B$ and the state of the complete system reads
\BE
\ket{\Psi_2}_{A,B}\equiv\hat U_{QFT}\hat U_{ph}\ket{\Psi_0}_{A,B} =\frac{1}{\sqrt{N}}\sum\limits_{n,\ell=0}^{N-1} {\cal W}_n^{(N)}(\ell) \;\ket{\ell}_A\ket{n}_B\label{def:psi_2_AB}
\EE
where   ${\cal W}_{n}^{(N)}(\ell)$ denotes the Gauss sum  defined in \eq{def:cal_W}.

This operation achieves two tasks: (i) the sum of quadratic phase terms is now independent of system $B$. For this reason, we are able to make a measurement on system $B$ leaving the sum of the quadratic phase terms in tact; and (ii) in addition to them a second phase term which is linear in $m$ arises. 

After a measurement on system $B$ with outcome $\ket{n_0}_B$, system $A$ is in the quantum state
\BE
\ket{\psi_3}_A \sim \sum\limits_{\ell=0}^{N-1}{\cal W}_{n_0}^{(N)}(\ell)\;\ket{\ell}_A.\label{eq:finite_entanglement}
\EE

The sum ${\cal W}_{n_0}^{(N)}$ is  equivalent to $G$ only for $n_0=0$. Nevertheless, ${\cal W}_n^{(N)}(\ell)$ shows properties which are similar to but not exactly the same  as $G(\ell,N)$. Therefore, we have to investigate now the influence of $n_0$ on ${\cal W}_{n_0}^{(N)}$.


\subsection{Probability distribution of system $A$\label{sec:investigation1} }

In this section, we discuss the probability distribution 
\BE
P_A^{(n_0)}(\ell,N)\equiv {\cal N}(n_0) |{\cal W}_{n_0}^{(N)}(\ell)|^2
\EE
 of system $A$, provided the measurement result of system $B$ is equal to $n_0$, and  analyze  its factorization properties. Here, $\mathcal{N}$ denotes a normalization constant.
Furthermore, we investigate how the measurement outcome $n_0$ of system $B$ influences these properties.  

From  \ref{app:P_n(l)} we recall the result
\BE 
|{\cal W}_{n_0}^{(N)}(\ell)|^2=\left\lbrace
\begin{array}{cl}
\frac{1}{N}&{\rm{if}}\;{\rm{ gcd}}(\ell,N)=1\\
\frac{p}{N}&{\rm{if}}\;{\rm{ gcd}}(\ell,N)=p{\rm{\; \& \;gcd}}(n_0,p)=p\\
0&{\rm{if}}\;{\rm{ gcd}}(\ell,N)=p{\rm{\; \&\; gcd}}(n_0,p)\neq p
\end{array}\right.,\label{eq:P(l)}
\EE
and recognize that there is a distinct difference between trial factors $\ell$ which share a common divisor with $N$ and trial factors which do not. Depending on whether  $n_0$ is (i) equal to zero, (ii) shares a common factor $p$ with $N$, or (iii) shares no common divisor with $N$, the probability for factors and their multiples is much higher than for other trial factors, or equal to zero. In any case, it is possible to distinguish between factors and nonfactors.

Now, we investigate the abilities of these three classes of probability distributions  to factor the number $N$.


\subsubsection{$n_0$ is equal to zero\label{subsec:n_0=0}}

\begin{figure}[ht]
\begin{center}
\includegraphics[width=0.65\columnwidth]{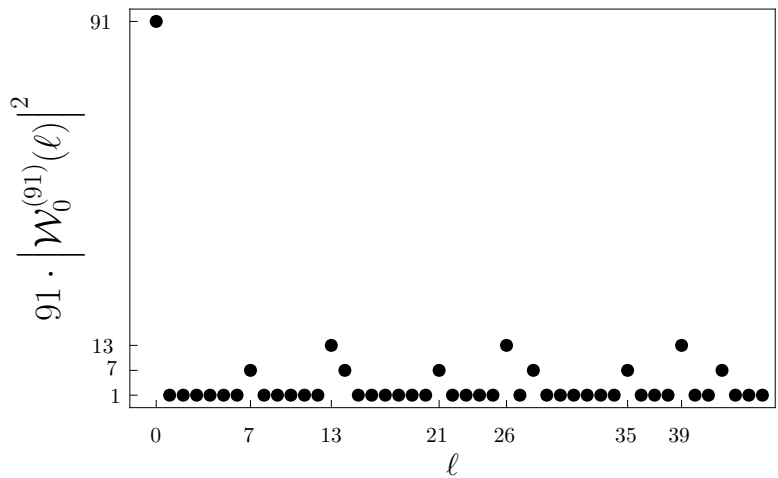}
\end{center}
\caption[Factorization of $N=91$ with $P_A^{(0)}(\ell,91)$]{Factorization of $N\;=\;91\;=\; 7\cdot 13$ with the help of the probability distribution $P_A^{(0)}(\ell;91)$to find the state $\ket{\ell}_A$ in system $A$ if we have measured before  the state $\ket{n_0=0}_B$ in system $B$. This probability distribution is proportional to $|{\cal W}_0^{(91)}|^2$. The probability for $\ell=0$ is $N=91$ times larger than for  trial factors which do not share a common divisor with $N$. For arguments $\ell$ which are multiples of a factor $p=7$ or $13$ the probability is $p=7$ or $13$ times larger, respectively. 
}
\label{fig:Ent_0_91}
\end{figure}

A special case occurs for $n_0=0$ where the probability distribution $P_A^{(0)}(\ell,N)$ is equal to the absolute value squared of the Gauss sum $G(\ell,N)$. It is the only case, where the probability $P_A^{(n_0)}(\ell=0,N)$ is nonzero. Indeed, here it is $N$ times larger than for trial factors, which do not share a common divisor with $N$. However, also  in this case the probability to find a multiple  of any factor $p$ of $N$ is $p$ times larger compared to arguments which do not share a common factor with $N$. It is for this reason that  the multiples of the factors $p=7$ and $13$  stand out in \fig{fig:Ent_0_91}. 

Important for the present discussion is not the probability for a given $\ell$ itself, but the probability ${\cal P}_A^{(0)}$ to find any multiple of  a factor. Now, we assume that $N$ contains only the two prime factors $p$ and $q=N/p$. In this case, there exist $N/p-1$ multiples of the factor $p$ with the probability $p/(4N-2p-2N/p+1)$ and  $p-1$ multiples of the factor $q=N/p$ with the probability $(N/p)/(4N-2p-2N/p+1)$. Therefore, ${\cal P}_A^{(0)}$ is given by
\BE
{\cal P}_A^{(0)}=\frac{2N-p-N/p}{2(2N-p-N/p)+1}.
\EE
For large integers $N$  we can neglect the term $+1$ in the denominator and arrive at the asymptotic behavior
\BE
{\cal P}_A^{(0)}\xrightarrow{}{N\rightarrow \infty}\frac{1}{2}.
\EE

As a consequence, the probability to find any multiple of a factor tends for large $N$ to $1/2$ independent of the prime factors. Therefore, the probability distribution $ P_A^{(0)}(\ell,N)$ is an excellent tool for factoring.


\subsubsection{$n_0$ and $N$ share a common divisor $p$\label{subsec:cd=p}}

\begin{figure}[ht]
\begin{center}
\includegraphics[width=0.65\columnwidth]{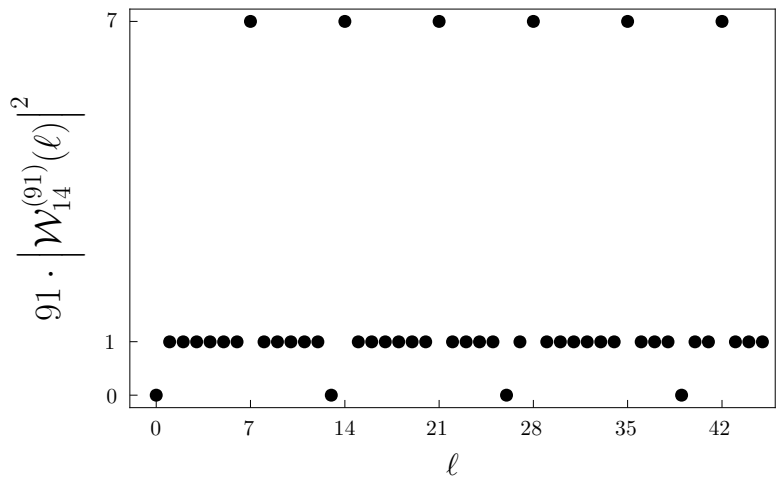}
\end{center}
\caption[Factorization of $N=91$ with $P_A^{(14)}(\ell;91)$]{Factorization of $N\;=\;91\;=\; 7\cdot 13$ with the help of the probability distribution $P_A^{(14)}(\ell;91)$ to find the state $\ket{\ell}_A$ in the system $A$  if we have measured  before the state $\ket{n_0=14}_B$. This probability is proportional to $|{\cal W}_{14}^{(91)}|^2$.  For $\ell$ being a multiple of $7$ the probability is seven times larger than for other trial factors, but for the multiples of the factor $13$, the probability vanishes, because $n_0=14$ is not a multiple of $13$.
}
\label{fig:Ent_14_91}
\end{figure}

If $n_0$ is a multiple of $p$ with $N=p \cdot q$ the probability to  find $\ell=k \cdot p$ is $p$ times larger than for other trial factors $\ell$. But the probability to measure a multiple of $q$ is equal to zero. This fact is clearly visible in \fig{fig:Ent_14_91} where we factor the number $N=91=7 \cdot 13$ with the help of $P_A^{(14)}(\ell,91)$. Because $n_0=14$ shares the common factor $7$ with $N=91$, all multiples of $7$ have a probability that is $7$ times larger  than arguments which do not share a common factor. In contrast, the probability to obtain any multiple of $13$ is still zero.

In order to derive the probability ${\cal P}_A^{(k\cdot p)}$ to find any multiple of a factor we note that there exist $N/p-1$ multiples of the factor $p$ with probability $p/(2N-2p-N/p+1)$ and arrive at
\BE
{\cal P}_A^{(k\cdot p)}=\frac{N-p}{2(N-p)-N/p+1}.
\EE
This function is  monotonically decreasing for $2\leq p \leq N/2$. Therefore, we get the smallest probability for the largest possible prime factor of $N$, which is $N/2$ and leads to
\BE
{\rm{min}}\left({\cal P}_A^{(k\cdot p)}\right)=\frac{N/2}{N-1}\xrightarrow{}{N\rightarrow \infty}\frac{1}{2},
\EE
which for large $N$ also tends to $1/2$. 

As a consequence, the case $n_0=k\cdot p$ displays a similar behavior as $n_0=0$: The probability $P_A^{(k \cdot p)}(\ell,N)$ is also an excellent tool for factoring.


\subsubsection{$n_0$ and $N$ do not share a common divisor\label{subsec:no_cd}}

\begin{figure}[ht]
\begin{center}
\includegraphics[width=0.65\columnwidth]{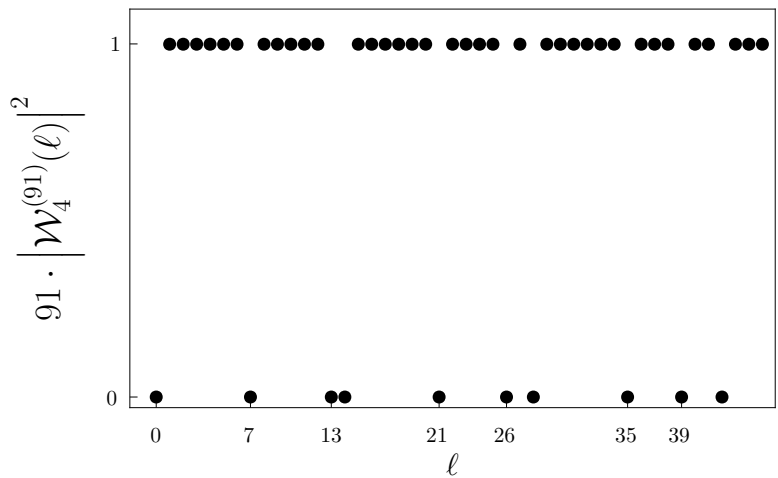}
\end{center}
\caption[Factorization of $N=91$ with $P_A^{(4)}(\ell;N)$]{Factorization of $N\;=\;91\;=\; 7\cdot 13$ with the help of the probability distribution $P_A^{(4)}(\ell;91)$ to find $\ket{\ell}_A$ in system $A$ if we have measured before  the state $\ket{n_0=4}_{B}$. The probability is proportional to $|{\cal W}_4^{(91)}|^2$. Clearly visible are the zeros when $\ell$ is a multiple of a factor such as $p=7$ or $13$. All other trial factors $\ell$ are equally probable.
}
\label{fig:Ent_4_91}
\end{figure}

As shown in Fig. \ref{fig:Ent_4_91} the probability $P_A^{(n_0\neq k\cdot p)}$ to find a multiple of a factor is equal to zero if $n_0$ and $N$ do not share a common divisor. As a consequence, it is not possible to deduce the factors of $N$ with a few measurements of the state of system $A$. Nevertheless, it should still be possible to extract the factors of $N$ from $P_A^{(n_0\neq k\cdot p)}$, although at the moment we do not know how to perform this task in an efficient way. However, there exist proposals that encoding information in the zeros  \cite{Fiddy1979} of a function is better than encoding them in the maxima. Therefore, we suspect that there may  exist an algorithm to obtain the information about the factors of $N$ from the zeros of $P_A^{(n_0\neq k\cdot p)}(\ell,N)$.

\subsection{Probability distribution of system $B$\label{sec:investigation2}}

In the preceding section we have found that the probability distribution of system $A$ depends crucially on the measurement outcome $n_0$ of system $B$. Depending on whether $n_0$ is a multiple of a factor of $N$ or not, system $A$ shows a different behavior. Therefore, it is essential to investigate the probability distribution of system $B$ for predicting the behavior of system $A$ which constitutes the topic of the present section.

With the help of the quantum state \eq{def:psi_2_AB} of the combined system the probability distribution 
\BE
P_B(n,N)\equiv\sum\limits_{\ell=0}^{N-1} \left|_A\bra{\ell} _B\bra{n}\ket{\Psi_2}_{A,B}\right|^2
\EE
for system $B$ is given by
\BE
P_B(n,N)= \frac{1}{N}\sum\limits_{\ell=0}^{N-1}\left|{\cal W}_n^{(N)}(\ell)\right|^2.\label{def:calP_N(n)}
\EE
From the explicit expression \eq{eq:P(l)} for $\left|{\cal W}_n^{(N)}(\ell)\right|$ we obtain the result
\BE
\fl P_B(n,N)=\frac{1}{N} \left\{ \left[\frac{1}{N}(N-p-q+1)+1\right] \delta_{n,0}+\frac{p}{N}(q-1)\delta_{\rm{gcd}(n,N), p}
+\frac{q}{N}(p-1)\delta_{\rm{gcd}(n,N),q}
\right\}\label{eq:P_B_ergebnis}
\EE
if $N$ is the product of the two prime numbers $p$ and $q$.

\begin{figure}[ht]
\begin{center}
\includegraphics[width=0.65\columnwidth]{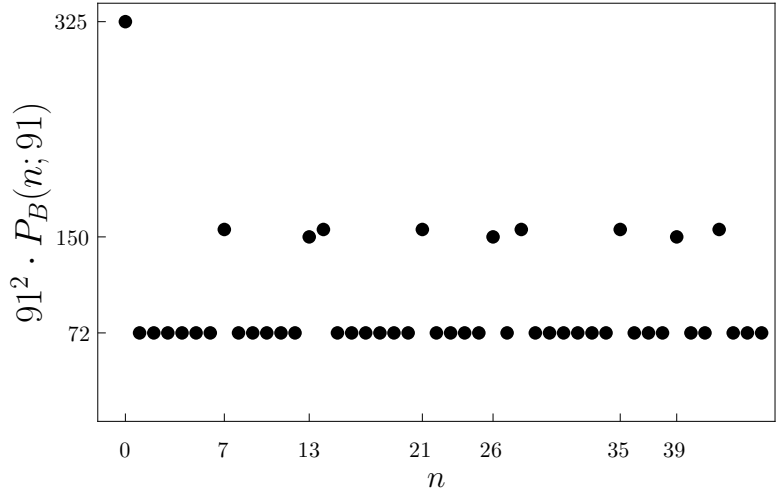}
\end{center}
\caption[Probability distribution $P_B(n;91)$ for $N=91$]{Probability  $P_B(n;91)$ for system $B$ to be in the state $\ket{n}_B$ for the example $N=91=7 \cdot 13$. For $n=0$ the probability is largest. For multiples of a factor the probability is enhanced compared to other trial factors.
}
\label{fig:Ent_n_91}
\end{figure}

An example for such a probability is depicted in \fig{fig:Ent_n_91}, where we show  $ P_B(n,N)$ for the example $N=91=7 \cdot 13$. The probability to find a multiple of the two factors $7$ and $13$ is larger than for other trial factors. 

However, we are not interested in the probability of a single $n$, but rather in  the probability ${\cal P}_B^{(0 {\rm{ \;or \;}} k\cdot p {\rm{ \;or \;}} k\cdot q)}$ that $n_0$ is equal to zero, or a multiple of a factor, because in these two cases it is possible to efficiently extract the information about the factors of $N$. Since ${\cal P}_B^{(0 {\rm{\; or \;}} k\cdot p {\rm{\; or \;}} k\cdot q)}$ is given by the sum over all probabilities $P_B(n,N)$ where $n$ falls into one of these cases that is
\BE
{\cal P}_B^{(0 {\rm{\; or\; }} k\cdot p {\rm{\; or\; }} k\cdot q)} \equiv  P_B(0,N) + \left(q-1 \right)  P_B(p,N)+ \left(p-1 \right) P_B(q,N)
\EE
we find with the help of \eq{eq:P_B_ergebnis} and $q=N/p$ the expression
\BE
{\cal P}_B^{(0 {\rm{\; or\; }} k\cdot p {\rm{ \;or \;}} k\cdot q)} = \frac{2}{Np}+\frac{2}{p}+\frac{2p}{N^2}+\frac{2p}{N}-\frac{p^2}{N^2}-\frac{4}{N}-\frac{1}{N^2}.
\EE

The probability ${\cal P}_B^{(0 {\rm{ \;or\; }} k\cdot p {\rm{\; or \;}} k\cdot q)}$ exhibits a minimum at $p=\sqrt{N}$, where it is given by
\BE
{\rm{min}}\left({\cal P}_M\right)= \frac{4N^{3/2}+4N^{1/2}-5N-1}{N^2}\xrightarrow{}{N\rightarrow \infty} \frac{1}{\sqrt{N}},
\EE
and it tends to zero for large $N$ as the inverse of a square root. As a consequence, it is very unlikely that $n_0$ is equal to zero, or a multiple of a factor but it is highly probable that $n_0$ shares no common divisor with $N$. Unfortunately, in this case a rapid factorization based on our algorithm is only possible if we find an efficient  way to extract the factors of $N$ from $P_A^{(n_0\neq k\cdot p)}$.


\subsection{Degree of entanglement\label{sec:degree_entanglement}}

In Sec.\ref{sec:algorithm} we have mentioned that the information about factors is contained in the entanglement of the two systems $A$ and $B$. Indeed,  \eq{eq:P(l)} suggests a strong correlation between the two systems. We now investigate the degree of entanglement\cite{Wang2006,Horodecki2009} with the help of the purity
\BE
\mu\equiv {\rm{Tr}}_i\left( \hat\rho_i^2\right)
\EE
of the  reduced density operator $\hat\rho_i$ with $i=A,B$. For a product state the purity is equal to unity. On the other hand, the state is maximally entangled if $\mu=1/D$, where $D$ denotes the dimension of the subsystem. We now derive an exact closed-form expression for $\mu$.

The reduced density operator $\hat \rho_A$ of system $A$ after the unitary transformation $\hat U_{ph}$ following from \eq{def:U_ph} reads
\BE
\rho_A=\frac{1}{N^2}\sum\limits_{\ell,\ell'=0}^{N-1}\sum\limits_{m=0}^{N-1}
\exp\left[2\pi \I m^2 \frac{\ell-\ell'}{N}\right]\ket{\ell}\bra{\ell'},
\EE
which is independent of applying a QFT on system $B$ or not. As a consequence, the purity  
\BE
\mu=\frac{1}{N^4}\sum\limits_{\ell,k=0}^{N-1} \sum\limits_{m,m'=0}^{N-1}\exp\left[2\pi \I (m^2-m'^2)\frac{\ell-k}{N}\right] 
\EE
of subsystem $A$  can be reduced to
\BE
\mu=\frac{1}{N^3}\sum\limits_{\ell=0}^{N-1}\left| \sum\limits_{m=0}^{N-1}\exp\left[2\pi \I m^2\frac{\ell}{N}\right] \right|^2
\EE
and is given by the sum 
\BE
\mu=\frac{1}{N^3}\sum\limits_{\ell=0}^{N-1}\left|G(\ell,N)\right|^2
\EE
of the standard Gauss sum $G(\ell,N)$ over all test factors $\ell$. Assuming that $N$ only contains the two prime factors $p$ and $q=N/p$, the purity can be written in a closed form which results from the following considerations.

For $\ell=0$ the standard Gauss sum is given by $|G(\ell=0,N)|^2=N^2$. Furthermore there exist $N/p-1$ multiples of $p$ which leads to the standard Gauss sum $|G(\ell=k\cdot p,N)|^2=pN$ and $p-1$ multiples of $N/p$ with $|G(\ell=k\cdot N/p,N)|^2=N^2/p$. For all other trial factors $\ell$, there exist $N-p-N/p+1$ of them, the standard Gauss sum is given by $|G(\ell)|^2=N$. Therefore, the closed form of the purity reads
\BE
\mu = \frac{1}{N^3}\left[  N^2+pN(\frac{N}{p}-1)+\frac{N^2}{p}(p-1)\right.\left.+N(N-p-\frac{N}{p}+1)\right],
\EE
that is
\BE
\mu=\frac{4N-2p-2N/p+1}{N^2}.
\EE
The purity is maximal for $p=\sqrt{N}$ and tends in this case for large $N$ to $ 4/N $. The maximal value of purity is equal to  the minimal degree of entanglement. Since the maximal purity is only $4/N$ and therefore very small, the two systems $A$ and $B$ are strongly entangled.


\subsection{Realization with qbits\label{sec:qbits}}

So far, we have not discussed the resources and the time necessary for our proposed algorithm. Both of them depend strongly on the underlying physical systems. For example, we can use  two light modes for the two systems $A$ and $B$ and the most appropriate states are photon number states. As a consequence, the energy  needed to display all states $\ket{\ell}_{A}$ and $\ket{m}_{B}$ would grow exponentially with the number of digits of $N$. 

Therefore, it is more efficient to run the algorithm with qbits. Here, the number $Q$ of qbits  scales linearly with the digits of $N$. However, the use of qbits changes our algorithm a little bit. For example, the initial state $\ket{\Psi_0}_{A,B}$ given by \eq{def:Psi_0}  now reads
\BE
\ket{\Psi'_0}_{A,B}\equiv \frac{1}{2^Q}\sum\limits_{m,\ell=0}^{2^Q-1}\ket{\ell}_{A}\ket{m}_{B}
\EE
with $N^2 < 2^Q$, that is the dimension of the system is not anymore given by the number $N$ to be factored. Moreover, this state can be easily prepared by applying a Hadamard-gate  to each single qbit whereas the state
$\ket{\Psi_0}_{A,B}$ is hard to create. 

In the unitary phase operator $\hat U_{ph}$ defined in \eq{def:U_ph} the number $N$ to be factored is encoded in an external variable which is independent of the dimension of the system, and can therefore be chosen arbitrarily. However, the QFT works now on a system of the size $2^Q$ and therefore the final state
\BE
\ket{\Psi'_2}_{A,B}\equiv \hat U_{QFT} \hat U_{ph}\ket{\Psi_0'}_{A,B}
\EE
 before the  measurement on system $B$ is given by
\BE
 \ket{\Psi'_2}_{A,B}= \frac{1}{2^{Q/2}}\sum\limits_{\ell,n=0}^{2^Q-1} \tilde{{\cal W}}_n^{(N)}(\ell,2^Q)\ket{\ell}_{A}\ket{n}_{B},
\EE
where we have introduced
\BE
\tilde{{\cal W}}_n^{(N)}(\ell,M)\equiv \frac{1}{M}\sum\limits_{m=0}^{M-1}\exp\left[2\pi\I \; (m^2\frac{\ell}{N}+m \frac{n}{M})\right].
\EE
We emphasize that this Gauss sum is a generalization of the Gauss sum ${\cal W}_n^{(N)}(\ell)$ defined by \eq{def:cal_W} due to the different denominators in the quadratic and linear phase. For reasons we have denoted this Gauss sum $\tilde{{\cal W}}_n^{(N)}(\ell,M)$ includes two arguments.

For the investigation of $\tilde{{\cal W}}_n^{(N)}$, we rewrite the summation index 
\BE
m\equiv sN+k
\EE
as a multiple $s$ of $N$ plus $k$ and find
\BE
\fl \tilde{{\cal W}}_n^{(N)}(\ell,2^Q)= \frac{1}{2^{Q}} \sum\limits_{s=0}^{M_N-1}\exp\left[2\pi \I s\frac{Nn}{2^Q}\right] \quad \sum\limits_{k=0}^{N-1}\exp\left[2\pi \I \left(k^2 \frac{\ell}{N}+k \frac{n}{2^Q}\right)\right]+R(l).
\EE
The remainder
\BE
R(l)\equiv \frac{1}{2^{Q}}\sum\limits_{k=0}^{r-1}\exp\left[2\pi \I  \frac{k^2\ell+k\cdot n}{N}\right]
\EE
with $MN+r=2^Q$ consists of less than $N$ terms, whereas the other part contains almost $N^2$ terms. As a consequence, we neglect $R$ and the probability $P'_B(n,N)$ to measure $n$ in system $B$ is approximately given by
\BE
\fl P'_B(n,N)\approx\frac{1}{2^{3Q}} \sum\limits_{\ell=0}^{2^Q-1}\left|\sum\limits_{k=0}^{N-1}\exp\left[2\pi \I \left(k^2 \frac{\ell}{N}+k \frac{n}{2^Q}\right)\right]\right|^2 \left|F\left(\frac{nN}{2^Q};M_N\right)\right|^2
\EE
where we have recalled the definition of $F\left(nN/2^Q;M_N\right)$ from \eq{def:F(n,m,Q)2}.

\begin{figure}[ht!]
\begin{center}
\includegraphics[width=0.65\columnwidth]{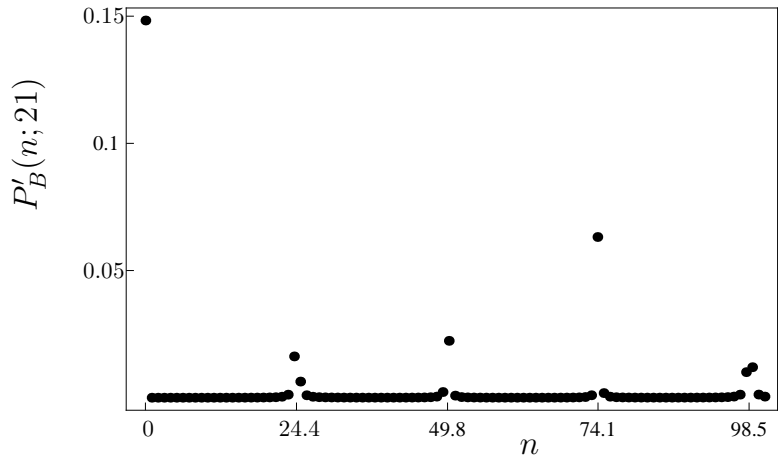}
\end{center}
\caption[Probability distribution $ P'_B(n;21)$ for $N=21$]{Probability  $P'_B(n;21)$ for system $B$ to be in the state $\ket{n}_B$ for the example $N=21=7 \cdot 3$ employing  $Q=9$ qbits for the algorithm. The probability is sharply peaked at multiples of $2^Q/N\approx 24.4$.
}
\label{fig:W_n_qbit}
\end{figure}

According to  \ref{chap:investigation_S}, the function  $F\left(nN/2^Q;M_N\right)$ is sharply peaked in the neighborhood of
\BE
n_N\equiv j \frac{2^Q}{N}+\delta_j\label{def:n_j}
\EE
with $|\delta_j| \leq 1/2$  if $N^2<2^Q$.  This behavior is depicted in  \fig{fig:W_n_qbit} for the example $N=21$ and $Q=9$. 

Since according to \ref{chap:investigation_S} the sum $F$ can be approximated by
\BE
F\left(\frac{nm}{2^Q};M_N\right)\approx \exp\left[\pi \I \delta_j\right] \frac{2}{\pi}M,
\EE
we can estimate the probability 
\BE
{\cal P}'_B\equiv \sum\limits_{n_N} P'_B(n_N,N)
\EE 
to find any $n_N$ defined  by \eq{def:n_j} by
\BE
{\cal P}'_B\approx \frac{4}{\pi^2}\frac{2^{2Q}}{N^2}\frac{1}{2^{3Q}} \sum\limits_{\ell=0}^{2^Q-1}\sum\limits_{j=0}^{N-1} \left| \sum\limits_{k=0}^{N-1}\exp\left[ 2\pi \I \left( k^2 \frac{\ell}{N}+k \frac{j}{N}\right)\right]\right|^2
\EE
that is
\BE
{\cal P}'_B\cong \frac{4}{\pi^2}\frac{1}{2^{Q}} \sum\limits_{\ell=0}^{2^Q-1}\sum\limits_{j=0}^{N-1}  | {\cal W}_j^{(N)}(\ell)|^2.\label{PBstrich}
\EE
By using \eq{eq:P(l)} we find
\BE
\sum\limits_{j=0}^{N-1}  | {\cal W}_j^{(N)}(\ell)|^2 = 1 \label{gleich_1}
\EE
by the following considerations: if ${\rm gcd}(\ell,N)=1$ then  $| {\cal W}_j^{(N)}(\ell)|^2 =1/N$ for all $j$ and \eq{gleich_1} follows at once; if ${\rm gcd}(\ell,N)=p$ then  $| {\cal W}_j^{(N)}(\ell)|^2 =p/N$ only for $j= k \cdot p$ and zero for all other $j$. Since in this case we have $q$ such terms we again obtain \eq{gleich_1}.

As a consequence of \eq{gleich_1}, the probability ${\cal P}'_B$  given by \eq{PBstrich} reduces to
\BE
{\cal P}'_B\cong \frac{4}{\pi^2}\frac{1}{2^{Q}} \sum\limits_{\ell=0}^{2^Q-1}1=  \frac{4}{\pi^2}
\EE
and system $A$ is with a probability greater than $40\%$ in the state
\BE
\ket{\psi_3'}_A \sim \sum\limits_{\ell=0}^{2^Q-1}{\cal W}_{j}^{(N)}(\ell)\;\ket{\ell}_A,
\EE
which is similar to the  state $\ket{\psi_3}_A$ given by \eq{eq:finite_entanglement} obtained by the algorithm   described in Sec.\ref{sec:algorithm}. However, it differs from it by the upper limit and by the fact that now the measurement outcome of system $B$ is not $n_0$ but the multiple $j$ of $2^Q/N$ defined by \eq{def:n_j}. Nevertheless, the properties of the states necessary to factor $N$ are the same.

\subsection{Discussion\label{sec:discussion}}

The analysis presented in this section was motivated by the idea to replace in the Shor algorithm the modular exponentiation by the standard Gauss sum. Indeed, the approach of Sec.\ref{chap:like_Shor} has lead us to  the problem how  to prepare the initial state containing the standard Gauss sum which we could solve in the present section by encoding the Gauss sum $ {\cal W}_n^{(N)}(\ell)$ into the probability amplitudes. However, this technique suffers again from the desease, that we arrive at  a state, where the factors of $N$ are encoded in the absence  of certain states. As a consequence, the search for an algorithm, which detects in an efficient way   the periodic appearance of missing  states is the most important task for the future of factorization with Gauss sums.

Moreover, we emphasize that by encoding the  Gauss sum in the probability amplitudes  we did not use the periodicity of the function itself, which was important in the Shor algorithm. The feature of the  Gauss sum central to an effective factorization scheme is the fact that  although there exist many more integers $\ell$  which are useless in factoring the number $N$, the product of their amount  times their probability is nearly equal to the amount of integers which do help us times their probability. This is the important difference to the truncated Gauss sum ${\cal A}_N^{(M)}$. Here, the ratio between the probability of  factors  to non-factors can be as small as $1:1/\sqrt{2}$. Furthermore, the truncated Gauss sum only exhibits maxima for the factors themselves and not for their multiples.


\section{Summary\label{sec:summary}}

So far the major drawback of Gauss sum factorization has been its lack of speed. Therefore,  we have combined in the present paper the Shor algorithm with the factorization with Gauss sums. Here, we have used the features of the absolute value $|G(\ell,N)|^2$ of the standard Gauss sum. Since  $G$ shows similar periodic properties as the function $f(\ell,N)=a^{\ell} \modu N$ which plays an important role in the Shor algorithm, we have replaced  $f$ by $g(\ell,N)\equiv |G(\ell,N)|^2/N$ and have investigated the resulting algorithm. We have shown that $f$ is not only special because  it is a periodic function, but also because there does not exist two arguments within one period which exhibit the same value of the function. This feature is the main difference to $g$, where nearly all arguments within one period lead to the same functional value. Therefore, we face the problem, that we have to distill the period of $g$ out of the zeros of a probability distribution instead of its maxima. Furthermore,  by replacing $f$ by $g$ the QFT is not necessary anymore.

Another challenge of our combination of Shor with Gauss sums, is the creation of the initial state 
$\ket{\Psi}_{A,B}$ defined by \eq{eq:like_shor_start}
because $g$ consists of a sum of complex numbers instead of integers. We have circumvented this problem by encoding the closely related  Gauss sum $  {\cal W}_n^{(N)}$ in the probability  amplitudes instead of the state. Furthermore, the number of terms in the standard Gauss sum grows exponentially with the number of digits of $N$ which makes it necessary to develop  implementation  strategies different from the ones which had been sucessful \cite{Experiments} with the truncated Gauss sum. Therefore, we have shown how to realize the Gauss sum ${\cal W}_n^{(N)}$ with the help of entanglement in an efficient way. Unfortunately, the resulting algorithm also sufferes from the problem that we need an efficient method to extract information from the zeros of a probability distribution.

In summary, we have investigated the similarities and differences of the Shor algorithm compared to Gauss sum factorization, which has lead us to a deeper understanding of both algorithms. Although, we have outlined a possibility for a fast Gauss sum factorization algorithm there is still the problem of the information being encoded in the zeros of the probability. As a consequence, the next challenge is to find an algorithm which performs this task  and paves the way for an efficient algorithm of Gauss sum factorization.


\section{Acknowledgment}
We thank M.~Bienert,  R.~Fickler, M.~Freyberger, F.~Haug,  M.~Ivanov, H.~Mack, W.~Merkel, M.~Mehring, E.~M.~Rasel, M.~Sadgrove, C.~Schaeff, F.~Straub and V.~Tamma for many fruitful discussions on this topic. This research was partially supported by the Max Planck Prize of WPS awarded by the Humboldt Foundation and the Max Planck Society.


\begin{appendix}


\section{Probabilities for the Shor  algorithm with Gauss sums}

In the present appendix we first derive an approximation for the sum
\BE
F\left(\alpha;M\right)\equiv \sum\limits_{k=0}^{M-1}\exp\left[2\pi\I k\;\alpha \right]\label{def:F(n,m,Q)}
\EE
at arguments $\alpha $ close to an integer $j$, that is $\alpha \equiv j+\delta_j/M$ with the upper bound $M$ and $|\delta_j|\leq 1/2$.
We then apply this approximate expression to calculate the probabilities $\tilde{\cal P}_A^{(p,p)}$ and $\tilde{\cal P}_A^{(p,N)}$ discussed in Sec. \ref{chap:like_Shor}.


\subsection{Approximate expression for $F(\alpha;M)$\label{chap:investigation_S}}

 We establish with the help of the geometric sum
\BE
\sum\limits_{k=0}^{M-1}q^k = \frac{1-q^{M}}{1-q}
\EE
 a closed form expression of $F$ which reads
\BE
F(\alpha;M)= \frac{1-\exp\left[2\pi\I\alpha M\right]}{1-\exp\left[2\pi\I\alpha\right] }.\label{F_closed_form}
\EE
By factoring out the phase factor $\exp\left[\I\pi\alpha M\right]$ in the numerator and $\exp\left[\I\pi\alpha\right]$ in the denominator we are able to rewrite \eq{F_closed_form} as
\BE
F(\alpha;M)= \exp[\I\pi \alpha (M-1) ] \quad\frac{\sin\left(\pi \alpha M\right)}{\sin\left(\pi\alpha\right)}
\EE
which is a ratio of two sine functions. This function displays maxima at integer arguments $\alpha=j$. For $\alpha =j+\delta_j/M$ we arrive at
\BE
F(j+\delta_j/M;M)\approx \exp[\I\pi \delta_j ] \quad\frac{\sin\left(\pi \delta_j \right)}{\sin\left(\pi\frac{\delta_j}{M}\right)}\label{eq:F(alpha)}.
\EE
Here, we have made use of the approximation $(M-1)/M\approx 1$ and  the fact, that $\sin(\pi(k+x))= (-1)^k\sin(\pi x)$ and $\exp[\I\pi k]=(-1)^k$ for integer $k$ and that for odd $j$ one of the two expressions $jM$ and $j(M-1)$ is even. 

\begin{figure}[ht!]
\begin{center}
\includegraphics[width=0.8\columnwidth]{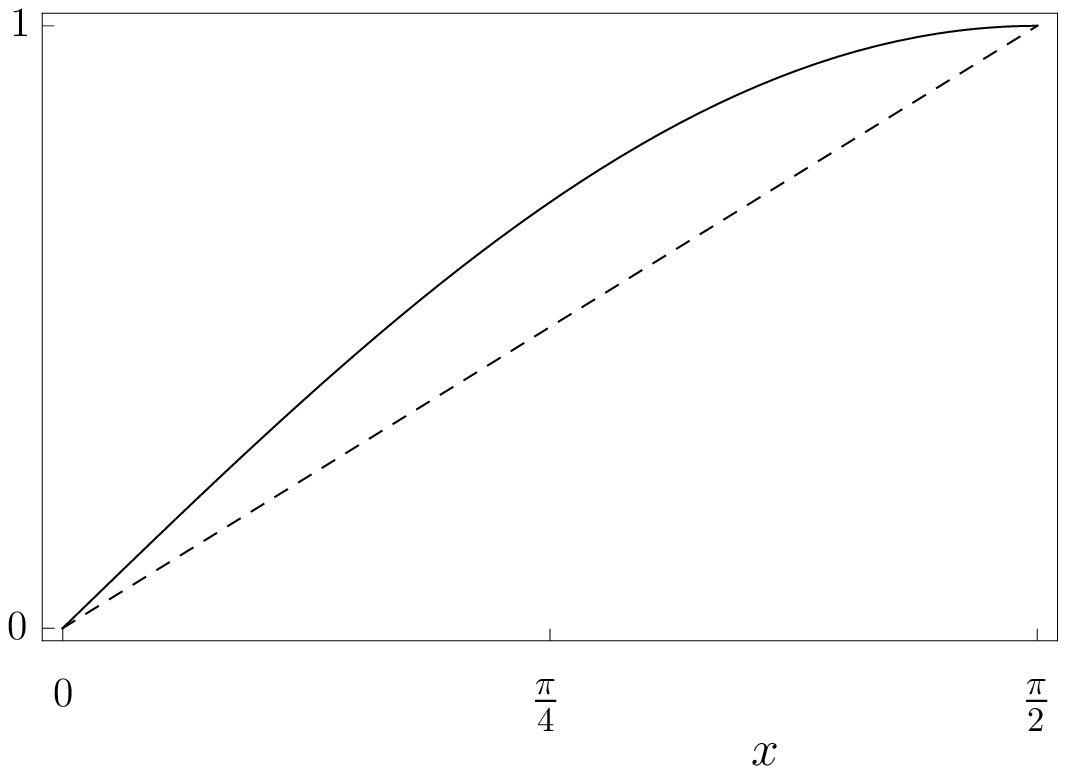}
\caption[Approximation of $\sin x$]{Graphical demonstration of the inequality $2x/\pi \leq \sin x$ for $0\leq x \leq \pi/2$. Here, the solid line depicts the function $\sin x$ whereas the dashed line represents $f(x)\equiv 2x/\pi$.}\label{fig:Approx_sinus}
\end{center}
\end{figure}

The argument $x\equiv\pi \delta_j$ of the sine function in the numerator  lies in the regime $0\leq  x \leq \pi/2$ and can therefore be approximated by
\BE
2\delta_j  \leq \sin\left(\pi \delta_j\right) 
\EE 
as we demonstrate graphically in \fig{fig:Approx_sinus}.

Furthermore, the sine function in the denominator of \eq{eq:F(alpha)} can be approximated by $\sin\left(\pi \delta_j/M\right)\approx \pi \delta_j/M$ because its argument $x= \pi \delta_j/M$ is much smaller than unity. Hence, we obtain
\BE
F(j+\delta_j/M;M)> \exp[\I\pi \delta_j ]\; \frac{2}{\pi}M\label{result_F}
\EE
as the final result.


\subsection{Calculation of probabilities $\tilde{\cal P}_A^{(p,p)}$ and $\tilde{\cal P}_A^{(p,N)}$\label{app:P_p}}

To estimate the probabilities $\tilde{\cal P}_A^{(p,p)}$  and $\tilde{\cal P}_A^{(p,N)}$ to find any multiple of $2^Q/p$ or of $2^Q/N$, if a measurement of system $B$ resulted in the factor $p$, we have to investigate the probability distribution
\BE
\tilde P_A^{(p)}(m;N)\equiv \frac{1}{2^Q(M_p-M_N)}\left|
\left[ F\left(\frac{pm}{2^Q};M_p\right) -F\left(\frac{Nm}{2^Q};M_N\right)
\right]\right|^2
\EE
for $m_p\equiv j \cdot 2^Q/p+\delta_j$ and $m_N\equiv j \cdot 2^Q/N+\delta_j$, with $M_N\equiv [2^Q/N]$ and $M_p\equiv [2^Q/p]$.

We first discuss the situation for $m_p$ and 
evaluate the function $F$ at the arguments
\BE
\frac{pm_p}{2^Q}= j+\delta_j \frac{p}{2^Q}\approx j+\frac{\delta_j}{M_p}
\EE
and
\BE
\frac{Nm_p}{2^Q}= qj+\delta_j \frac{N}{2^Q}\approx qj+\frac{\delta_j}{M_N}
\EE
using the estimate \eq{result_F} 
\BE
F(j+\delta_j/M;M)> \exp[\I\pi \delta_j ]\; \frac{2}{\pi}M
\EE
for $|\delta_j \leq 1/2|$. 

As a consequence,  the probability to find a state $\ket{m}$ with $m$ being close to  a multiple $j$ of $2^Q/p$ reads
\BE
\tilde P_A^{(p)}(m_p;N )> \frac{1}{2^Q(M_p-M_N)}\frac{4}{\pi^2}\left(M_p^2+M_N^2-2 M_p M_N \right)
\EE
which reduces with the help of the binomial formula $x^2+y^2-2xy = (x-y)^2$ to  
\BE
\tilde P_A^{(p)}(m_p;N )>\frac{M_p-M_N}{2^Q}\frac{4}{\pi^2}.
\EE

When we recall that $M_N\approx 2^Q/N$ and $M_p\approx 2^Q/p$ we obtain the final result
\BE
\tilde P_A^{(p)}(m_p;N )> 0.4 \frac{N-p}{Np}
\EE
with the approximation $4/\pi^2> 0.4$. 

Since there exist $p$ different values of $m_p$ the total probability $\tilde P_A^{(p,p)}$ to find any multiple of $2^Q/p$ is given by
\BE
\tilde {\cal P}_A^{(p,p)}> 0.4 \frac{N-p}{N}
\EE
which tends towards $0.4$ for large $N$ and  prime factors $p \leq \sqrt{N}$.

We now calculate the probability  $P_A^{(p)}(m_N;N)$ to find  $m_N= j\cdot  2^Q/N+\delta_j$ which are not multiple of $m_p$. At these arguments the  sum $F\left(pm_N/2^Q;M_p\right)$ is close to zero and therefore can be neglected. As a consequence, we can approximate $\tilde P_A^{(p)}(m_N;N)$  by 
\BE
\tilde P_A^{(p)}(m_N;N)>0.4\frac{ M_N^2}{2^Q(M_p-M_N)}\approx 0.4 \frac{p}{N(N-p)}. 
\EE
Since there exist $N-p$ different values of $m_N$ the total probability $\tilde {\cal P}_A^{(p,N)}$ to find any multiple of $2^Q/N$ is given by
\BE
\tilde {\cal P}_A^{(p,N)} > 0.4 \frac{p}{N}
\EE
which tends towards zero for large $N$ and  prime factors $p \leq \sqrt{N}$.


\subsection{Calculation of probability $\tilde P_A^{(1)}$\label{app:tilde_P_2}}

Similar to the calculation in the last section, we now evaluate the probability $\tilde P_A^{(1)}$ to find any multiple $p$ or $q$ if the measurement of system $B$ resulted in unity. For this task, we first estimate the probability distribution
\begin{eqnarray}
\fl \tilde P_A^{(1)}(m;N)&\equiv& \left|_A\bra{m}\hat U_{QFT}\ket{\psi^{(1)}}_A\right|^2 \nonumber \\
&=&\frac{1}{2^Q(2^Q-M_p-M_q+M_N)} 
\left|F\left(\frac{m}{2^Q};2^Q\right)-F\left(\frac{pm}{2^Q};M_p\right)\right.\nonumber \\ &&\left.-F\left(\frac{qm}{2^Q};M_q\right)+F\left(\frac{Nm}{2^Q};M_N\right)\right|^2
\end{eqnarray}
for $m_p\equiv j \cdot 2^Q/p+\delta_j$, $m_q\equiv j \cdot 2^Q/q+\delta_j$ and $m_N\equiv j \cdot 2^Q/N+\delta_j$ following from \eq{eq:QFT_psi_1_A} with $M_q\equiv [2^Q/q]$.

The first term given by $F$ is equal to zero for all $m \neq 0$. The second term leads to peaks at multiples of $2^Q/N$, the third to peaks at multiples of $2^Q/p$ and the last to peaks at multiples of $2^Q/q$. In this case, we get information about the factors of $N$ only from the peaks at multiples of $2^Q/p$ and $2^Q/q$. 

As a result we obtain the probability
\BE
\tilde P_A^{(1)}(m_p;N)>0.4\; \frac{(M_p-M_N)^2}{2^Q(2^Q-M_p-M_q+M_N)}
\EE
to find $m_p$. Here, we have taken into account that $m \approx j\cdot 2^Q/p$ is also an integer multiple of $2^Q/N$. With the help of the approximations $M_N\approx 2^Q/N, M_p\approx 2^Q/p$ and $M_q\approx 2^Q/q$ we get the final result
\BE
\tilde P_A^{(1)}(m_p;N)>0.4\;\frac{(q-1)^2}{N(N-q-p+1)}.
\EE
Similar, we obtain
\BE
\fl \tilde P_A^{(1)}(m_q;N)>0.4\; \frac{(M_q-M_N)^2}{2^Q(2^Q-M_p-M_q+M_N)}=0.4\;\frac{(p-1)^2}{N(N-q-p+1)}
\EE
for the probability to find $m \approx j\cdot 2^Q/q$.

For $\tilde P_A^{(1)}(m_N,N)$ only the term $F\left(Nm/2^Q;M_N\right)$ is non-vanishing which leads to
\BE
\fl \tilde P_A^{(1)}(m_N;N)>0.4\; \frac{M_N^2}{2^Q(2^Q-M_p-M_q+M_N)}=0.4\;\frac{1}{N(N-p-q+1)}.
\EE 
As a consequence, we arrive at  the total probability
\BE
\fl \tilde {\cal P}_A^{(1, p {\rm{ \;or \;}}q)}= p \tilde P_A^{(1)}(m_p;N)+  q \tilde P_A^{(1)}(m_p;N) >0.4\frac{Nq+Np+q+p-4N}{N(N-q-p+1)}
\EE
 to find any multiple of a factor $p$ or $q$.


\section{Probabilities for the superposition algorithm\label{app:P_n(l)}}

In this appendix, we calculate the probability  
\BE
P_A^{(n_0)}(\ell,N)={\cal N}(n_0) |{\cal W}_{n_0}^{(N)}(\ell)|^2
\EE
to measure $\ell$ in system $A$ if the measurement result of system $B$ was $n_0$. Here, $\mathcal{N}$ is a normalization constant and $\mathcal{W}_{n_0}^{(N)}$ is defined as
\BE
{\cal W}_n^{(N)}(\ell)\equiv \frac{1}{N}\sum\limits_{m=0}^{N-1}\exp\left[2\pi\I \; (m^2\frac{\ell}{N}+m \frac{n}{N})\right].
\EE

Since  $P_A^{(n_0)}(\ell,N)$ is proportional to ${\cal W}$ we take advantage of the result 
\BE
|{\mathcal W}_{n_0}^{(N)}(\ell)| = \sqrt{\frac{1}{N}}\label{eq:betrag_W}
\EE
from Ref.\cite{Schleich2001}. Here, $N$ is odd and  $\ell$ does not share a common divisor with $N$. We are only interested in factoring odd numbers $N$. If we have to factor an even number, we can divide it  by two repeatedly until we arrive at an odd number.

If  $\ell$ and $N$ share a common divisor $p$ we have to eliminate it  before we are allowed to apply \eq{eq:betrag_W}. Assuming that
\BE
\ell\;=\;k \cdot p,\quad N\;=\;q \cdot p
\EE
with $k$ and $q$ coprime, the sum ${\mathcal W}_{n_0}^{(N)}$ reduces to 
\BE
{\cal W}_{n_0}^{(q\cdot p)}(k \cdot p)\equiv \frac{1}{N}\sum\limits_{m=0}^{N-1}\exp\left[2\pi\I \; (m^2\frac{k}{q}+m \frac{n_0}{N})\right].
\EE
Now, the quadratic phase is periodic with period $q$ and not with period $N$. Therefore, it is useful to rewrite the summation index $m$ as
\BE
m\;=\;r\cdot q +s
\EE 
and cast the Gauss sum
\BE
{\cal W}_{n_0}^{(q \cdot p)}(k \cdot p)=\sum\limits_{r=0}^{p-1} \sum\limits_{s=0}^{q-1} \exp\left[2\pi\I\left(\frac{k}{q}s^2+\frac{n_0}{N}(rq+s)\right) \right]
={\cal F}(n_0,p) \cdot {\cal W}_{n_0}^{(q)}(k ) 
\EE
into the form of a product of two sums, where 
\BE
{\cal F}(n_0,p) \equiv \sum\limits_{r=0}^{p-1}\exp\left[2\pi\I\; \frac{n_0}{p}r\right]
\EE
points out the role of $n_0$: it is equal to $p$ if it is a multiple of $p$. Otherwise, it vanishes. 

We now  apply \eq{eq:betrag_W} to evaluate  $|{\cal W}_{n_0}^{(q)}(k )|$ and find
\BE
|{\cal W}_{n_0}^{(N)}(\ell)|^2=\left\lbrace
\begin{array}{cl}
\frac{1}{N}&{\rm{\;if\; gcd}}(\ell,N)=1\\
\frac{p}{N}&{\rm{\;if \;gcd}}(\ell,N)=p{\rm{ \& gcd}}(n_0,p)=p\\
0&{\rm{\;if\; gcd}}(\ell,N)=p{\rm{\; \& \;gcd}}(n_0,p)\neq p
\end{array}\right. 
\EE
We emphasize   that here we define ${\rm{gcd}}(0,N)\equiv N$. 

The normalization  constant $ {\cal N}$ follows from the condition
\BE
\sum\limits_{\ell=0}^{N-1}P_A^{(n_0)}(\ell,N)=1
\EE
and reads
\BE
|{\cal N}(n_0)|^2 \equiv \left\lbrace \begin{array}{cl}
\frac{N}{4N-2p-2q+1} & {\rm{for\; }} n_0=0\\
\frac{N}{2N-2p-q+1} & {\rm{for\; gcd}}(n_0,N)=p\\
\frac{N}{N-p-q+1}& {\rm{else}}
\end{array}\right.
\EE
assuming $N$ contains only the two prime factors $p$ and $q$.

\end{appendix}

\section*{References}
\bibliographystyle{bst}

\end{document}